# United classification of cosmic gamma-ray bursts and their counterparts

Alexander Kuznetsov

**Abstract** United classification of gamma-ray bursts and their counterparts is established on the basis of measured characteristics: photon energy $E$ and emission duration $T$. The founded interrelation between the mentioned characteristics of events consists in that, as the energy increases, the duration decreases (and vice versa). The given interrelation reflects the nature of the phenomenon and forms the $E - T$ diagram, which represents a natural classification of all observed events in the energy range from $10^9$ to $10^{-6}$ eV and in the corresponding interval of durations from about $10^{-2}$ up to $10^8$ s. The proposed classification results in the consequences, which are principal for the theory and practical study of the phenomenon.

**Keywords** Gamma rays: bursts

## 1 Introduction

Intensive investigations of cosmic gamma-ray bursts (GRBs) have been carried out beginning with 1973. They were observed for the first time (Klebesadel et al. 1973) as pulse gamma-radiation in the energy range from 0.3 to 2 MeV with the burst duration of $T \sim 10$ s.

Gamma-ray bursts and concomitant radiation (GRBs counterparts) are considered here as the unique phenomenon, which is observed in the energy range from $\sim$ GeV to $10^{-6}$ eV. As the energy decreases, the glow duration increases from $10^{-2}$ to $\sim 10^8$ s. GRBs are measured, as a rule, in the energy range of 20 keV – 2 MeV with duration of $10^{-2} - 10^3$ s. The GRB counterpart includes X-ray ($10^{-1} - 10^2$ keV), optical, UV, NIR (0.56 – 6.2 eV), and radio glow at frequencies of 1.4 – 350 GHz ($6\times10^{-6} - 10^{-3}$ eV), as well as the high-energy radiation – from 10 MeV to $\sim$ GeV. At present we cannot restrict the observed radiation in energy either from above or from below.

Directions of arrival of GRB and its counterpart are isotropically distributed over the celestial sphere. They coincide within the limits of modern accuracy of measurement, up to one angular second. This suggests that the same object represents a source of both GRB and its counterpart. In this paper, for simplicity, the words "GRB and its counterpart" may be replaced by the words " the source glow" or simply "the glow".

Numerous works establish the unity of the physical process generating the electromagnetic radiation throughout the observed energy range, for example (Piro et al. 1998). This unity is confirmed by the compatibility of light curves and spectra. The results of some works are presented briefly below.

The unified spectrum describes the glow for GRB 990123 in the energy range from 10 keV to 10 MeV (Briggs et al. 1999).

The unity of light curves within a very wide energy range is observed for GRB 970508. The gamma-ray burst smoothly, without delay, transfers into X-ray radiation, which is overlapped with optical one, and the latter one – with a radio glow. The $E \sim T$ interrelation is emphasized: GRB $\sim$ 10 s, X-rays and optical radiation $\sim 10^5$ s (Piro et al. 1998). A considerable energy release in various ranges, comparable with GRB energy, supposes that the same process is responsible for GRB and for concomitant radiation. The duration of radio glow for GRB 970508 is traced up to $\sim 4\times10^7$ s (Frail et al. 2000b). The relationship between the energy of photons and duration of glow (decrease of energy – increase of duration) is well revealed at transition from one separated energy range to another.

The glow is observed within a very wide energy range, which can be either continuous, as in the case of GRB 970508, or with discontinuities (Briggs et al. 1999; Connors & Hueter 1998). Since the subdivision into ranges of energy is rather conventional (because it is substantially determined by the measurement technique) in the general case one should expect discontinuities – the dark spots in any interval of energies and durations.

The appearance of the robotic optical telescopes resulted in measuring the minimum delay of optical radiation relative to the beginning of GRB ($T_0$), which equals 22 s for GRB 990123 (Akerlof et al. 1999) and 21.8 s for GRB 050801 (Rykoff et al. 2006). On the basis of these measurements the delay of X-ray radiation relative to GRB can be estimated as ~ sec. The comparison of GRB 930131 measurements by the COMPTEL instrument (0.7 – 20 MeV) and BATSE (20 keV – 1.2 MeV) shows the 0.1 s delay of GRB beginning recorded by BATSE (Ryan et al. 1994). For one of most powerful GRB 830801.b, the flux from which in the range of 0.03 – 7.5 MeV is estimated as $\geq 2 \cdot 10^3$ erg/cm$^2$, the hard radiation of 1450 – 2500 keV appears with 1-s outstripping of soft radiation of 165 – 255 keV (Kuznetsov et al. 1986). Because the nature does not know what are the gamma-, optical radiations, etc., it is worth to consider the observed time delay (outstripping) between corresponding values of radiation energies. For example, the delay of optical radiation relative to the GRB corresponds to the delay of radiation with energy of eV relative to the energy of $10^6$ eV. In comparing the time of beginning of hard radiation relative to softer one the outstripping is observed and in the opposite case the delay takes place. To date, no delay is observed in the time of high-energy radiation relative to softer one.

This note results in supposition that the separation of observed radiation into GRB and afterglow is erroneous. In this connection the terms "counterpart" or "concomitant radiation" are used in the paper instead of "afterglow".

The fact of existence of insignificant energy-dependent delay (outstripping), obviously, reflects the peculiarity of the emission process. One should note that the mentioned delay (outstripping) and the duration of emission represent two different characteristics of events, which depend on the photon energy.

The $E \sim T$ interrelation is studied on the basis of observational data in distinguished energy ranges: high-energy gamma radiation, gamma-ray bursts, X-ray, optical and radio glow. Proceeding from the general $E \sim T$ dependence, at which the emission duration increases from a fraction of second to $10^8$ s, as the energy decreases from $10^9$ down to $10^{-6}$ eV, the extreme values of the energy-duration are determined for each of distinguished energy ranges: $E_{max}$, $T_{min}$ and $E_{min}$, $T_{max}$. Each point on plots presents the event with measured coordinates $E$, $T$. So, we find that all observed events of the distinguished range have the $E$, $T$ values in the limits from $E_{max}$, $T_{min}$ to $E_{min}$, $T_{max}$, thus establishing the orderliness of all observed events. Connecting by the line the extreme points, separately, $E_{max}$, $T_{min}$ and $E_{min}$, $T_{max}$ of all distinguished ranges, we obtain 2 boundary curves, between which the coordinates of all phenomenon's events are concluded. The knowledge of coordinates of extreme $E - T$ values makes it possible to define the type of curves determining the boundaries of the phenomenon, as well as the type of curves connecting the extreme points of each distinguished range.

The work is based on the observational data obtained in the experiments of Compton Gamma-Ray Observatory (*CGRO*) BATSE, OSSE, COMPTEL and EGRET, as well as in the *BeppoSAX*, KONUS-WIND, PHEBUS, *Ulysses*, *Swift* and other experiments, working, mainly, simultaneously with CGRO. How far possible, the work has used the data obtained by the instruments and telescopes with extremely high sensitivity in order to provide correspondence of obtained results to the modern level of accuracy of physical measurements. The emission duration dependence on instrument's sensitivity was taken into account. So, the observation of the UVOT (*Swift*) instrument in the absence of observations by ground-based telescopes have not actually used in the paper. The durations of some short GBS measured in the KONUS – WIND experiment [Mazets et al. 2004] were corrected to durations measured by BATSE instrument.

**2 Interrelation between photon energy and glow duration**

Gamma-ray bursts, X-ray, optical and radio emissions represent relatively narrow and well-studied ranges of energy (the glow domains). High-energy gamma-ray bursts represent scantily explored, but, probably, most interesting events for the phenomenon origin. For obtaining boundary lines of the $E - T$ diagram in each of aforementioned ranges two pairs of boundary values - two points $E_{max}$, $T_{min}$ and $E_{min}$, $T_{max}$, which are just presented in Fig. 1, are determined below.

2.1 High-energy gamma-ray bursts

This high-energy gamma radiation is presented by bursts measured in the energy range from some MeV up to GeV. Only a few events are known, for which the photons with the energy up to ~ GeV were measured. Unfortunately, the measurements in X-ray, optical and radio ranges could not be performed for these events; so, the observations of associated glow throughout the energy range of $10^{-6} - 10^9$ eV are absent.

The GRB 910503 spectrum was measured by BATSE and COMPTEL instruments in the energy ranges of 0.03 – 3.1 MeV (Band et al. 1993), 0.6 – 10 MeV (Hanlon et al. 1994), respectively. The EGRET Total Absorption Shower Counter (TASC) and the EGRET Spark chamber (SC) recorded single photons with energy of 50 – 200 MeV (Schneid et al. 1992). The duration of the burst measured by BATSE and COMPTEL equals about 60 s. The EGRET (SC) instrument has recorded 9 photons in the range of 50 – 233 MeV during 15 s after GRB beginning ($T_0$). In this case 3 photons of 9 were not coincided in direction of arrival with GRB, and, hence, they were not considered to be associated with the given GRB. In total, the EGRET (SC) recorded 46 photons, including aforementioned 6 photons, during 505 s. The photons were observed before the GRB beginning as well. This case, where the EGRET (TASC) and EGRET (SC) have detected radiation in the range up to ~ 200 MeV simultaneously, allows us to make some interesting comparisons. The GRB light curve consists of 2 peaks of duration about 2 seconds each, they are separated by the time of ~ 50 s, but the second peak being not observed at energies higher than 7 MeV. Six high-energy photons with the energies 49, 101, 233, 89, 55, 93 MeV and corresponding time intervals between photons 0.36, 0.42, 0.17, 0.37, 0.34 s coincide in time with the first peak in the light curve recorded by BATSE. The comparison of ~ 2 s duration of the burst recorded by the EGRET (TASC) with energy higher than 30 MeV with the < 2 s duration of recording 6 photons with energies from ~ 50 to 233 MeV in the EGRET (SC) indicates that the real burst duration for energy of ~ 50 MeV will be close to 2 s. However, the duration of high-energy gamma-ray emission of 100 – 200 MeV will be ~ 0.4 s. This example demonstrates the features of recording high-energy gamma radiation and determining its characteristics. In connection with the energy-duration dependence high-energy photons observed during 505 s raise the doubts: whether they really represent emission from the GRB source?

The GRB 910601 was observed by the BATSE, OSSE, COMPTEL and EGRET (TASC) instruments in the ranges of 0.01 – 3.1 MeV, 0.06 – 9.5 MeV, 0.6 – 23 MeV, 1 – 52 MeV (Schaefer et al. 1998), respectively. The GRB duration, measured by BATSE, equals ~ 45 s, the duration of glow, observed by COMPTEL, is estimated as 33 s (Hanlon et al. 1994). The difference in duration indicates, first of all, existing of the dependence: decrease of energy – increase of glow duration.

The GRB 930131 was recorded by the BATSE, COMPTEL and EGRET (TASC) instruments in the energy ranges of 0.02 – 1.18 MeV, 0.7 – 20 MeV (Ryan et al., 1994), 1 – 180 MeV (Sommer et al. 1994), respectively. In addition, the EGRET (SC) has recorded 18 photons with energy higher than 30 MeV for the time of about 100 s. The first 16 photons of these 18 were detected during the time of 25 s after $T_0$; two of these photons with energy of about 1 GeV have time 2.4 and 25 s after $T_0$ (Sommer et al. 1994). The duration of a burst, measured by BATSE, equals 19 s, whereas according to the EGRET (TASC) and COMPTEL data, this burst duration does not exceed 2 s. Within the limits of these 2 s the EGRET (SC) recorded 3 photons with energies of ~ 80, 30 and 460 MeV, which, apparently, can be agreed with the range of energies observed by EGRET (TASC). In this case there is also in doubt the large duration of a burst (of about 100 s) measured by the EGRET (SC) in the energy range of 40 MeV – 1 GeV, as compared to the duration of ~ 2 s measured by other instruments in the energy range of 1 – 180 MeV. There is no correlation in time for high-energy photons of ~ 1 GeV with the GRB light curve measured up to the energy of 180 MeV.

The GRB 940217.2 was measured by the BATSE, Ulysses, COMPTEL, EGRET (TASC) and EGRET (SC) instruments in the energy ranges of 0.03 – 3 MeV, 25 – 150 keV, 0.3 – 30 MeV, 1 – 200 MeV, 40 MeV – 18 GeV, respectively (Hurley et al. 1994a). The GRB duration, measured by BATSE, Ulysses and EGRET (TASC) equals 180 s, COMPTEL – 162 s. During 162 s the burst emission of high-energy photons up to 4.4 GeV were recorded by the EGRET instrument (Hurley et al. 1994b, c). On the subsequent orbit of *CGRO*, 90 minutes later, EGRET (SC) detected 10 photons, one with energy of 18 GeV – within the 20 minutes time interval from the burst region. EGRET (TASC) also detected during that interval radiation in the energy bands 1.1 – 6.1 MeV and 6.1 – 26.2 MeV, respectively (Hurley et al., 1994c). However the COMPTEL during that time observing the same source did not detect any significant signal in the 0.3 – 10 MeV range (Winkler et al. 1995). The spectrum measured by EGRET (SC) is not a high-energy extrapolation of the EGRET (TASC) MeV data (Hurley et al. 1994c).

Single measurements allow to state the reliable detection of high-energy gamma radiation up to 200 MeV, based on its temporal and spectral compatibility with GRB. Actually, the value of 200 MeV represents an observable upper limit of glow in energy for today. In order to consider the radiation of ≥ GeV as a part of a glow, it is necessary to have confirmation of its relationship with the glow source that is also noticed in paper (Connors & Hueter 1998).

For the $E - T$ dependence Fig. 1 presents three points of high-energy radiation with approximate coordinates $E$, $T$: 50 MeV, 2 s and 200 MeV, 0.4 s (GRB 910503), and also ~ 20 MeV, 7 s (GRBs 940206 and 960124) from PHEBUS experiment (Tkachenko et al. 1998, 2002).

2.2 Gamma-ray bursts

The GRBs are considered, which are observed in the gamma-ray range $E_\gamma = 0.2 - 2$ MeV with duration of ~ $10^3 - 10^{-2}$ s.

For gamma-bursts of $\Delta T \sim 10^{-2} - 10^3$ s and $\Delta E \sim 2000 - 200$ keV the maximum duration should correspond to the minimum energy of ~ 200 keV. The GRB 020410 measured by the KONUS instrument in the energy range of 15 – 200 keV has duration $T \sim 2300$ s. This GRB really has maximum energy of ~ 200 keV (Nicastro et al. 2004) and occupies, as noted, the position between typical GRBs and X-ray rich events.

For GRB 900813, measured by the PHEBUS instrument, the maximum energy equals $2000 \pm 600$ keV and the burst duration equals $0.02 \pm 0.01$ s (Terekhov et al. 1994), so that the boundary point on the $E_{max}$, $T_{min}$ line has coordinates 2 MeV, 0.02 s. Two boundary points: $E_{max}$, $T_{min}$ – 2000 keV, 0.02 s and $E_{min}$, $T_{max}$ – 200 keV, 2300 s are designated in Fig. 1 by black circles and by digits 1 and 2.

The PHEBUS experiment, which records GRBs in the energy range of 0.1 – 30 MeV, well defines the boundary point on the $E_{max}$, $T_{min}$ line – 500 keV, 0.07 s, (GRB 931008), (Tkachenko et al. 1998). In Fig. 1 this point is marked by digit 3.

2.3 X-ray glow

The X-ray concomitant radiation was investigated in the range of $E$ = 2 – 10 keV. According to the *BeppoSAX* data (De Pasquale et al. 2006a) this energy range corresponds to $T$ ranging from 7 s (GRB 971227) to $6 \times 10^5$ s (GRB 981226), which provides two boundary points: $E_{max}$, $T_{min}$ – 10 keV, 7 s and $E_{min}$, $T_{max}$ – 2 keV, $6 \times 10^5$ s. These points are presented in Fig. 1 by digits 5 and 6a.

The observed maximum duration of X-ray radiation in the range of 2 – 10 keV equals ~ $6 \times 10^5$ s, which more than 100 times exceeds the GRB maximum duration.

The study of X-ray glow in the interval of $E$ = 0.2 – 10 keV was carried out by the X-Ray Telescope (XRT) in the *Swift* observatory. The maximum glow duration, which equals $8.6 \times 10^6$ s, is marked in the event concomitant to GRB 060729 (Grupe et al. 2007), which must correspond to the energy of 0.2 keV. This point $E_{min}$, $T_{max}$ – 0.2 keV, $8.6 \times 10^6$ s is designated in Fig. 1 by digit 6b.

2.4 Optical glow

The concomitant optical glow observed in the energy interval 0.56 – 6.2 eV has duration from several days to ~ a year. The maximum of glow duration corresponding to GRB 970508 measured in the R-band (2.07 eV) equals $2.3 \times 10^7$ s (Panaitescue & Kumar 2004). This duration approximately 40 times exceeds the maximum duration of glow in the X-ray range 2 – 10 keV.

The minimum glow duration, which corresponds to GRB 050416A, was observed in the M2-band (5.6 eV) is $7 \times 10^4$ s (Holland et al. 2007). The boundary points $E_{max}$, $T_{min}$ – 5.6 eV, $7 \times 10^4$ s and $E_{min}$, $T_{max}$ – 2.07 eV, $2.3 \times 10^7$ s are marked in Fig. 1 by digits 7 and 8.

2.5 Radio glow

The durations of concomitant radio emission, which is measured, as a rule, at frequencies of 1.4 – 350 GHz ($6 \times 10^{-6}$ – $1.5 \times 10^{-3}$ eV), equal from several days to ~ 800 days. The minimum glow duration measured at frequency of 350 GHz (GRB000926) is 3.3 days (Smith et al. 2001). The maximum glow durations, measured at frequencies of 250 GHz (GRB 010222), 8.5 GHz (GRB 000210) and 1.4 GHz (GRB 980425) equal to date 73 days (Frail et al. 2002), 610 and 790 days (Frail et al. 2003), respectively. The measured maximum duration of $7 \times 10^7$ s exceeds the maximum duration of optical glow as low as three times. In Fig. 1 digits 9, 10, 12 and 14 denote aforementioned four points with coordinates $1.45 \times 10^{-3}$ eV, $2.8 \times 10^5$ s; $10^{-3}$ eV, $6 \times 10^6$ s; $3.5 \times 10^{-5}$ eV, $5 \times 10^7$ s and $6 \times 10^{-6}$ eV, $7 \times 10^7$ s.

3 Discussion of energy – duration diagram

All points $E_{max}$, $T_{min}$ (1, 3, 5, 7) and $E_{min}$, $T_{max}$ (2, 4, 6a, 6b, 8), and others determined above are shown in Fig. 1. The connection of point's 3, 7 and point's 2 and (50 MeV, 2s) forms two nearly parallel boundary lines AB and CD. One should emphasize that points 1, 3, 7 and 2, 6a, 6b, and others, which represent measurements at edges of separated ranges, automatically designate boundary lines. The events cannot be observed out of the boundary lines by the definition. The boundary lines in the log scale are determined by the dependences $E = k_1 T$, which allow to calculate the glow duration $T$ from the given energy $E$ and vice versa. The slope angle of AB line determined by points 3, 7 is, approximately, equal $\alpha_1 = -40°$; the slope angle for CD line, calculated by points 2 and (50 MeV, 2s) is equal $\alpha_2 = -38°$. The best fit of the data determining the AB and CD lines is given by the power function $y = ab / (b + x)$, the plot of which represents in the linear scale a curve of hyperbolic type. The apexes of AB and CD curves of hyperbolic type have coordinates 1 keV, 109 s and 30 keV, 26125 s, respectively. Any point of the diagram can be considered as a result of a single measurement, but this point also belongs to the set of lines – cross-sections of the diagram. The dotted lines (1 – 2, 1 – 4, etc.) connecting boundary points of measurements in separated ranges are also determined by the dependence of $E = k_2 T$, which provide calculation of one of quantities $E$, $T$ from the known other quantity. These lines represent cross-sections of the diagram. Cross-sections of the diagram by lines $E$ = const ($E_{max} = E_{min}$) we use in tracing the change of the E – T dependence on the high-energy edge of the phenomenon.

The slope angle of the 1–2 line, that describes the range of GRBs, is $\beta_1 = -11°$. This angle changes with a change of the range of measurements: it will increase with expanding the range – descending along the CD line or ascending along the AB line. So, as the range of GRBs energies expands from 2000 – 200 keV to 2000 – 80 keV point 2 displaces to point 4 (Fig. 1), and the new range will be presented by the 1–4 line with corresponding change of the slope angle. In the linear scale these lines are the curves of hyperbolic type as well. The line (for example, 1–2) shifted parallel to itself by a small quantity will also represent GRBs but now in the other range of the energies-durations.

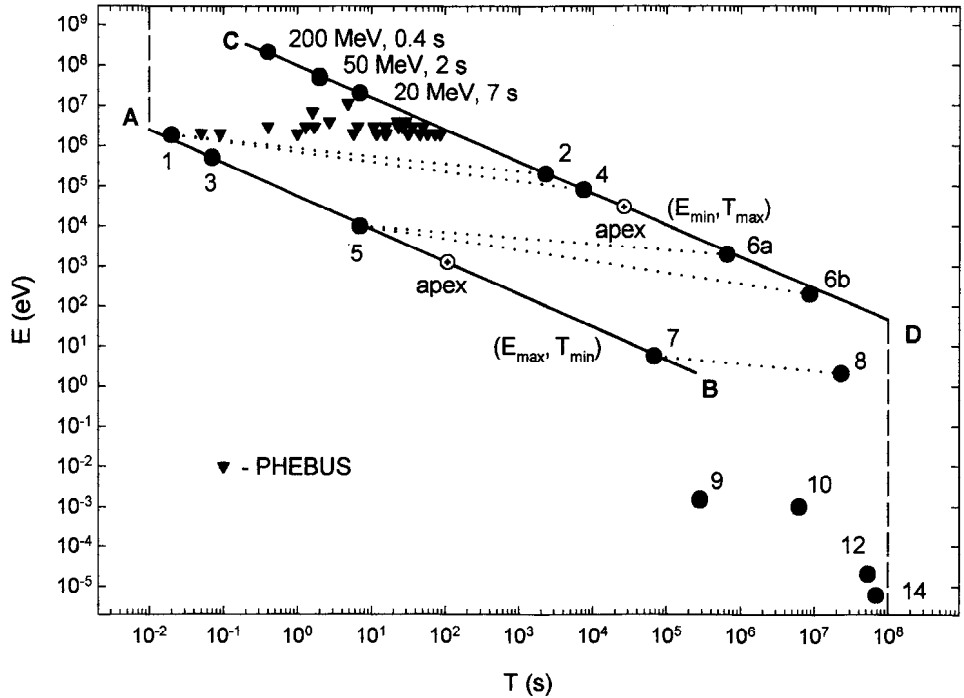

**Fig. 1.** Diagram reflects the dependence between the photon energy and glow duration. The basic part of the diagram consists of the band restricted by lines AB and CD, which is limited in the energy, 2 MeV $\geq E \geq$ 100 eV, appreciatively. The data designated by *triangles* display the existence of the HE edge of the phenomenon. The observed limitations of the events in the glow duration are marked on the diagram by *vertical dashed lines*.

**4  More precise definition of phenomenon boundaries**

The use of a considerable number of events allows to more precisely determine the $E$ - $T$ dependence and, accordingly, the shape of a diagram. The Fig.2 is obtained, actually, by simple mapping of the majority of accessible events in the energy range $6\times10^{-6} \div 2 \times10^8$ eV on the plot in the $E, T$ coordinates. Below the position of these events in the diagram and some features are discussed.

KONUS-WIND events. There are in all about 130 GRBs from the catalogue of short events (Mazets et al. 2004). The catalogue contains the full information on GRBs: the time history, duration, the energy interval, in which the burst was observed, as well as the energy spectrum. These events are presented on the diagram of Fig. 2 in terms of glow duration and maximum energy of photons. They designate the boundary of the phenomenon; their extrapolation over the upper level allows to make a rough estimation of the maximum possible photon energy of ~ 200 – 300 MeV. It should be noted that the KONUS GRBs, as well as the events of other experiments having nearly identical characteristics (repeated events), can be not shown in the Fig. 2.

The GRB 010420a recorded in the energy range of 15 – 1000 кeV has minimum duration among 130 events, which equals $T_{90} = 0.01$ s. This very weak burst, for which, obviously, the energy spectrum is absent due to this reason, is well seen in Fig. 2 – it escapes the limits of the diagram. The GRB 900813 (the PHEBUS experiment) has duration close to 0.01 s. The mentioned GRBs are, obviously, compatible within the limits of errors in energy and duration.

In the majority of events the durations are measured to a good accuracy – of about several percents. The energy of photons in the optical and radio ranges is measured at fixed frequencies within narrow bands, so that the given measurements can be considered to be monoenergetic ones. Such point-like measurements of $E, T$ are ideal for mapping the events in Fig. 2 and for obtaining the diagram. The measurement of energy of GRB and X-ray events is carried out within some interval $\Delta E = E_2 - E_1$, however. The lower threshold $E_1$ depends, as a rule, from the instrument sensitivity and is fixed. The measured duration of events corresponds to the observed maximum energy $E_2$.

BATSE 1 events. For more than 50 gamma-ray bursts detected for the first year of the BATSE experiment were calculated spectra (Band et al. 1993). Actually, these bursts represent long GRBs with the lower energy threshold of 10 – 30 кeV and the upper limit of 1200 – 4300 кeV. As seen from Fig. 2, all events are very well incorporated in the diagram in the duration and maximum energy of the measured spectrum and supplement the KONUS experiment data.

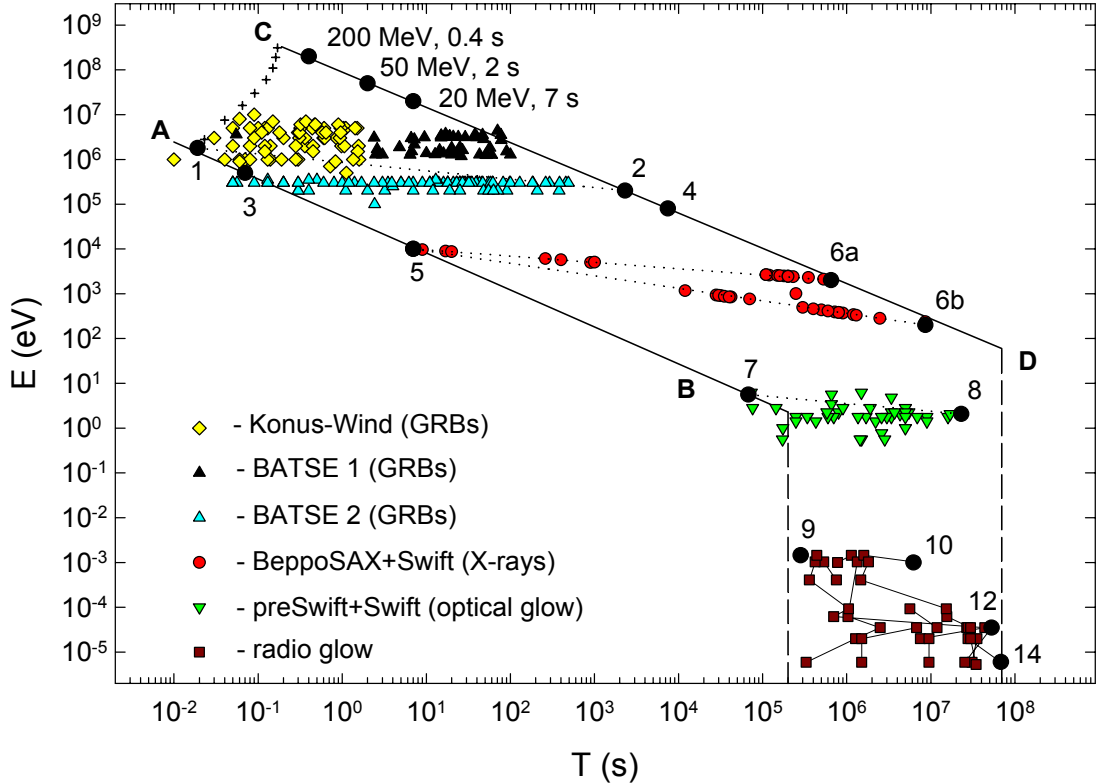

**Fig. 2.** *E - T* dependence changes throughout the range of events, making weaker from the HE to the LE edge of the phenomenon. The line designated by *crosses* is the supposed boundary at the HE edge. The slope angle of the event band between the lines AB and CD is about minus 40°. The smooth, gradual decreasing of the slope angle from –40° up to minus 90° at the LE edge shows the weakening of the *E – T* dependence up to its full disappearance, that corresponds to absence of any events.

BATSE 2 events. This set includes about 300 GRBs from the BATSE catalogue (Paciesas et al. 2000). All these events were measured during 1991 – 2000 in the BATSE experiment by the Large Area Detector in three channels: 25 – 60 кeV, 60 – 110 кeV and 110 – 300 кeV, i.e., totally, in the energy range of 25 – 300 кeV. Knowing that the maximum energy of photons in these GRBs does not exceed 300 кeV, one can perform the diagram cross section by the line *E* = 300 кeV. This cross-section presented in Fig. 2 would correspond to the true one, if all GRBs had the maximum energy close to 300 кeV. The real scattering of maximum energies is enclosed in the range of 110 – 300 кeV, however. This implies that, actually, the cross-section will have the band of about 200 кeV in width between the lines of *E* = 110 кeV and the line of *E* = 300 кeV. This band is partly shown by the events lying under the basic cross-section on the line of *E* = 200 кeV. The figure contains one purely X-ray event GRB 910822, in which the photon energy does not exceed 100 кeV. The events GRB 930725.2, GRB 950130, GRB 960710 and GRB 970201.2 escape the diagram limits to the left side. The light curves of these short events are presented in the catalogue by the line without indications to the presence of the time structure. In total, one can count more than 2 tens of such events from the BATSE catalogue, including a wider range of energies. If the internal structure is really absent in the considered events, they represent the false GRBs - some kind of artefact, most likely.

X-ray glow (*BeppoSAX + Swift*). All events, presented in Fig. 2, have been observed in the *BeppoSAX* – (2 – 10 кeV) and *Swift* – X-Ray Telescope (0.2 – 10 кeV) experiments.

*BeppoSAX*: The characteristics of the X-ray glow concomitant to gamma-bursts 960720, 970111, 970228, 970402, 970508, 971214, 971227, 980109, 980326, 980329.1, 980515, 980519, 980613, 990123, 010222 and 011211, are taken from (De Pasquale et al. 2006a). The X-ray event corresponding to GRB 981018 was considered (in 't Zand et al. 2003), GRB 981226 – (Frontera 2004), GRB 020410 – (Nicastro et al. 2004).

*Swift*: The data are presented for the following GRBs: 041223 (Burrows et al. 2005a); 050126 and 050219A (Goad et al., 2006); 050128 (Campana et al. 2005); 050215B (Levan et al. 2006); 050315 (Vaughan et al. 2006); 050319 (Cusumano et al. 2006); 050326 (Moretti et al. 2006); 050401 (De Pasquale et al. 2006b); 050408 (Capalbi et al. 2007); 050603 (Grupe et al. 2006); 050721 (Antonelli et al. 2006); 050406, 050421, 050502B, 050607, 050724, 050730, 050904 (Burrows et al., 2005b); 060729 (Grupe et al., 2007), 061007 (Schady et al., 2006). All remarks concerning BATSE 2 gamma-bursts position on the diagram relate to aforementioned X-ray events as well. These events are shown in the Fig. 2 in accordance with measured durations and calculated energies supposing their position along the 5 – 6a line (*BeppoSAX*) and the 5 – 6b line (*Swift*). In fact, the *BeppoSAX* events will be distributed in the energy band enclosed between 10 кeV and 2 кeV, and the *Swift* events – between 10 кeV and 0.2 кeV.

Optical glow (pre *Swift* + *Swift*). Pre *Swift* data include the events corresponding to the following GRBs: 970508 R-band, 000301c R-band (Panaitescu & Kumar 2004); 970228 R, 971214 R (Diercks et al. 1998); 980519 R, 991216 R, 000926 R, 010222 R, 011121 R, 011211 R, 020405 R, 030226 R (Zeh et al. 2006); 990123 KR (Kulkarni et al. 1999a); 990123 BVI, 990510 VRI (Holland et al. 2000); 000911 BVRIJH-bands (Mazetti et al. 2005).

*Swift* data include events corresponding to GRBs 041223 RK and 050124 KJ (Berger et al. 2005); 050215b K (Levan et al. 2006); 050416A W2M2BV (Holland et al. 2007); 050721 BRI (Antonelli et al. 2006); 060729 W2M2W1UBV (Grupe et al. 2007).

In Fig. 2 these events are reflected by more than 40 points representing the glow in the energy range from NIR to UV: 0.6 – 6.2 eV. The optical glow measurements are carried out within very narrow energy intervals – in fact, they represent monoenergetic measurements. This circumstance favorably discriminates these measurements from previous measurements in the gamma and X-ray ranges.

The glow in the radio range. The events are observed, as a rule, at frequencies of 1.43, 4.86, 8.46, 15, 22.5, 100, 250 and 350 GHz corresponding to the photon energy interval from $5.9 \times 10^{-6}$ eV to $1.45 \times 10^{-3}$ eV. The observations of radio, as well as of optical glow, represent monoenergetic measurements. The reliable coordinates of events, which are obtained as a result of such measurements, allow to correct the form of the diagram corresponding to the E - T dependence.

Figure 2 presents the glow durations concomitant to the following GRBs observed at separated frequencies: GRB 970508: 1.43, 4.86, 8.46 GHz (Frail et al. 2000b); 980329: 250, 350 GHz (Yost et al. 2002); 980425: 1.38, 4.80, 8.64 GHz, 000210: 1.43, 8.46 GHz and 000301c: 8.46 GHz (Frail et al. 2003); 990123: 1.43, 4.86, 8.46, 15, 350 GHz (Galama et al. 1999; Kulkarni et al. 1999b); 991208: 100, 250 GHz (Galama et al. 2003); 991216: 1.43, 4.86, 8.46 GHz (Frail et al. 2000a); 000301c: 4.86, 15, 22.5, 100, 250, 350 GHz (Berger et al. 2000); 000926: 350 GHz (Smith et al. 2001); 010222: 250, 350 GHz (Frail et al. 2002); 030329: 1.28, 4.86, 8.46, 15, 22.5, 100, 250 GHz (Resmi et al. 2005; Berger et al. 2003; Sheth et al. 2003); 031203: 1.43, 4.86, 8.46, 22.5 GHz (Soderberg et al. 2004).

The events observed at various frequencies and corresponding to concrete GRB in the Fig. 2 are linked by a solid line.

## 5 Summary

5.1 Estimation of phenomenon limits

One of consequences of this work is the possibility of estimating the limiting photon energies and glow durations.

The limitations in the emission duration are known from the experiments. The GRBs are not observed with duration lower than 0.01 s, which is reflected in Fig. 1 by the vertical dashed line $T = 0.01$ s. The cross-sections of the diagram by lines $E = 2$, 3, and 4 MeV and 2 events GRB 951014 (7 MeV, 1.6 s), GRB 920723 (11 MeV, 4.8 s) designated by triangles, are presented in Fig. 1 according to the PHEBUS data (Terekhov et al. 1994, 1995; Tkachenko et al. 1998, 2002). For the $E = 2$ MeV cross-section the glow durations $T$ have the values from $T_{min} = 0.02$ to $T_{max} = 86$ s, for the $E = 3$ MeV cross-section – from 0.4 to 50 s, and for the $E = 4$ MeV cross-section – from 2.7 to 30 s. These data indicate sharp narrowing of the band of events toward the boundary CD line. The reality of narrowing into a single point, eventually, is confirmed by the fact that photons with energy of 200, 50 and 20 MeV are located near the CD line. In the Fig. 2 the curve designated by crosses shows the supposed boundary of the phenomenon at the HE edge. The shape of this curve is assumed on the basis of the above-mentioned PHEBUS data. The all these features allow to suppose the existence of the upper limit of glow in the energy $E < 1$ GeV. The possibility of emission of photons with energy of 1 - 20 GeV within the phenomenon framework, apparently, should be considered separately, as some exception to the rule.

The LE radiation measured in the radio range of 1.4 – 350 GHz is not observed with duration greater than $\sim 10^8$ s, which is shown in Fig. 1 by the vertical dashed line $T = 10^8$ s. The position of optical and radio events in the Fig. 2 suggests a smooth, gradual decreasing of the slope angle of the event band from – 40° to – 90°. The latter slope angle – 90° is indicated in Fig. 2 by vertical dashed lines. This weakening of the $E - T$ dependence begins at the level of radiation with the energy of about some tens eV. Obviously, the limit of phenomenon in the energy at the LE edge and the corresponding limit in duration will be determined by full disappearance of the given dependence, which is equivalent to the absence of events at all. It is assumed that as the events approach the line closing the boundary at the LE edge, the $E - T$ dependence is absent at all. The measurements at the LE edge allow to suppose the existence of the lower limit of glow in the energy $E \sim 10^{-6}$ eV.

5.2 Nature of phenomenon

The dependences of photon energy and time run slowdown on the gravitation potential follow from the theory of gravitation – the general relativity theory (GR). So, the energy of a photon being in the gravitational field can be presented, in some approximation, as a sum consisting of the free photon energy h$\nu$ and its negative potential energy h$\nu$ $\times \varphi_N/c^2$: $E = h\nu(1 + \varphi_N/c^2)$, where $\varphi_N$ is the Newtonian gravitational potential (Vladimirov et al. 1984). This total energy is conserved as the photon moves from point 1 to point 2: $h\nu_1(1 + \varphi_N(1)/c^2) = h\nu_2(1 + \varphi_N(2)/c^2)$. Thus, some specific value of photon energy can be attributed to the value of gravitational potential at the given point.

Taking into account that the mentioned dependence is conspicuous in the strong gravitational field, one may suppose that the observed glow takes place in the field of a black hole. Then phenomenon is the black hole's glow. Also one can consider that the unique $E - T$ interrelation has the only possible explanation based on participation of a strong gravitational field in the process of emission of GRBs and their counterparts. The GRBs and their counterparts represent a completely new kind of electromagnetic radiation, in which the photon energy and the glow duration are synchronized. This relativistic phenomenon should accord completely to GR concepts. Obviously, its observed features (not only the energy and duration of glow) can provide a quantitative test for the contemporary theory of gravitation.

### 5.3 Conclusions

This paper shows the existence of such an interrelation between the photon energy $E$ and the emission duration $T$ where any fixed range of energy has the corresponding, specific time interval of durations. The dependence between the energy of photons $E$ and the duration of glow $T$ can be geometrically presented by the second order surface formed by the curves of hyperbolic type. The $E - T$ diagram represents the unique classification of all measured (and non-measured) events and shows a possibility of the unique catalog creation.

The law of correspondence of basic event characteristics is the manifestation of some law of nature, whose knowledge can be useful in preparation of experiments and allows, in general, to predict the expected results. So, from the positions of the short GRBs (Mazets et al. 2004) and long GRBs (Band et al. 1993) in Fig.2 one may propose that, starting from the energy ~ 50 MeV, the long GRBs will not be observed. Also the diagram shows that, starting from the energy ~ 10 keV and lower, the short events are not observed.

Obviously, the $E - T$ dependence underlies of such observed effects as the time evolution of the GRB spectrum and the GRB duration distribution. From the diagram (Fig. 2) follows that the parameters of this distribution will be defined, in general, by the energy range, in which the GRBs are measured.

The subordination of radiation characteristics $E$ and $T$ to a single law leads to the conclusion on the existence of a limited class of standard, nearly identical sources of the observed phenomenon.

The same physical process, as it is supposed in some works, generates the GRBs and their counterparts. The $E - T$ diagram confirms the unique origin of the phenomenon – the glow of a source. Gamma-ray bursts are only a part of this glow and nothing more. The glow is possible in any energy–duration range, both continuous or with discontinuities. Characteristic, probably, is the glow within a wide range of energies–durations.

**Acknowledgements**. Author thanks the anonymous referee for the constructive comments.